\documentclass[crop]{CSLB}

\jname{International Journal of Astrobiology}%

\begin{document}

\supertitle{Research Paper}

\title[Distinguishing multicellular life on exoplanets by testing Earth as an exoplanet]{Distinguishing multicellular life on exoplanets by testing Earth as an exoplanet}

\author[Doughty et al.]{Christopher~E.~Doughty$^{1}$, Andrew~Abraham$^{1}$, James~Windsor$^{2}$, Michael~Mommert$^{3}$, Michael~Gowanlock$^{1}$, Tyler~Robinson$^{2}$, and David~Trilling$^{2}$}

\address{\add{1}{School of Informatics, Computing, and Cyber Systems, Northern Arizona University, Flagstaff, AZ. 86011, USA}, \add{2}{Department of Astronomy and Planetary Science, Northern Arizona University, Flagstaff, AZ. 86011, USA} and \add{3}{University of St. Gallen, Institute of Computer Science, Rosenbergstrasse 30, 9000 St. Gallen, Switzerland}}

\corres{\name{Christopher~E.~Doughty} \email{chris.doughty@nau.edu}}

\begin{abstract}
Can multicellular life be distinguished from single cellular life on an exoplanet?  We hypothesize that abundant upright photosynthetic multicellular life (trees) will cast shadows at high sun angles that will distinguish them from single cellular life and test this using Earth as an exoplanet.  We first test the concept using Unmanned Arial Vehicles (UAVs) at a replica moon landing site near Flagstaff, Arizona and show trees have both a distinctive reflectance signature (red edge) and geometric signature (shadows at high sun angles) that can distinguish them from replica moon craters.  Next, we calculate reflectance signatures for Earth at several phase angles with POLDER (Polarization and Directionality of Earth's reflectance) satellite directional reflectance measurements and then reduce Earth to a single pixel.  We compare Earth to other planetary bodies (Mars, the Moon, Venus, and Uranus) and hypothesize that Earth’s directional reflectance will be between strongly backscattering rocky bodies with no weathering (like Mars and the Moon) and cloudy bodies with more isotropic scattering (like Venus and Uranus). Our modelling results put Earth in line with strongly backscattering Mars, while our empirical results put Earth in line with more isotropic scattering Venus.  We identify potential weaknesses in both the modelled and empirical results and suggest additional steps to determine whether this technique could distinguish upright multicellular life on exoplanets. 
\end{abstract}


\selfcitation{xx xxxx xxxx}

\maketitle

\section{Introduction}

Recently, a 1.3 Earth mass planet only $\sim$4 light years from Earth was found within the habitable zone of the red dwarf Proxima Centauri \citep{Anglada-Escude2016ACentauri}. According to the NASA Exoplanet Archive, by July 23 of 2020, 4197 exoplanets, have been confirmed  (https://exoplanetarchive.ipac.caltech.edu/), including one in the habitable zone with water vapor in its atmosphere \citep{Tsiaras2019WaterB}. Do these exoplanets have life and if so, what type of life might it be?   A number of techniques have been proposed to test whether life exists on exoplanets and many of these are summarized in recent reviews \citep{Schwieterman2018ExoplanetLife,Catling2018ExoplanetAssessment}.  The goal of all this is, of course, to be able to use next generation astronomical facilities to detect life on the recently discovered exoplanets (see review by~\citep{Fujii2018ExoplanetProspects}) ~\citep{Fujii2018ExoplanetProspects}.  However, such reviews have missed a critical stage – distinguishing an exoplanet with single cellular life from that of multicellular life.  Some have hypothesized that single cellular life may be abundant in the universe, but multicellular life may be rare \citep{Brownlee2000RareEarth}.  We clearly need a technique to distinguish between the two types of life.

Since photosynthesis could be abundant in the universe, what techniques, for example, could we use to distinguish the change between land covered with abundant terrestrial single-celled photosynthetic organisms like those in the Precambrian \citep{Kenny2001StablePrecambrian} and the rise of multicellular life, like the land plants that occupied Earth from the Mid-Ordovician (490–430 million years ago) to today \citep{Graham2000TheRadiation}?  Previous work has proposed that the most abundant multicellular life on an exoplanet would likely be vertical photosynthetic organisms – trees \citep{Doughty2010DetectingPlanets}.  The need to transport water and nutrients and competition for light in multicellular photosynthetic organisms has led to the tree-like structure on Earth characterized by hierarchical branching networks \citep{Brown2000ScalingBiology,West1997ABiology}.   In fact, the ``tree shape'' evolved independently many times throughout Earth’s history likely as a consequence of the previously mentioned biomechanical and evolutionary constraints \citep{Donoghue2005KeyPhylogeny}. Such biomechanical constraints combined with Darwinian evolution will also make tree-like photosynthetic structures the most abundant evidence of multicellular life on exoplanets.

\Fpagebreak %need this for the IJA template at the end of the first page

Earth has more than 3 trillion trees \citep{Crowther2015MappingScale}, each with a vertical structure that casts shadows differently than objects on a lifeless planet with weather and climate.  Almost all trees are at a 90$^\circ$ angle to the ground while less than 1 percent of the surface of the Earth has with a slope greater than 45$^\circ$ \citep{Hall2005CharacterizationData}.  This is simply because weather and climate, which are thought to be necessary on any planet capable of sustaining multicellular life \citep{Kasting2003EvolutionPlanet} will erode much abiotic topography over time.  For instance, one study suggested a lifeless planet with weather will be very similar to Earth topologically \citep{Dietrich2006TheLife}.   Therefore, shadows at certain sun angles may be indicative of multicellular life, but could we detect them on an exoplanet?

Earth Scientists know a great deal about tree shadows because to accurately estimate terrestrial reflectance (with, for example, Landsat or MODIS (Moderate Resolution Imaging Spectroradiometer) satellite data) shadows at different sun angles must be removed.  Therefore, a great deal of effort has been put into developing a quantitative framework to predict shadows at different sun angles.  This framework, called the bidirectional reflectance distribution function (BRDF), is the change in observed reflectance with changing view angle or illumination direction \citep{Schaepman-Strub2006ReflectanceStudies}.  Forests seen from different sensor sun angles have predictable differences in reflectance \citep{Breon2002AnalysisSpace,Breon2006SpaceborneDistributions,Li1992Geometric-OpticalShadowing,Wolf2010AllometricMeasurements}.  Previous work used a semi-empirical BRDF model \citep{Bacour2005VariabilityPOLDER,Maignan2004BidirectionalSpot} at the global scale to explore whether, in theory, Earth with vegetation would have different albedo at different sensor sun angles versus an Earth without vegetation \citep{Doughty2010DetectingPlanets}.  They found that even if the entire planetary albedo were rendered to a single pixel, the rate of increase of albedo as a planet approaches full illumination would be comparatively greater on a vegetated planet than on a non-vegetated planet.  It was hypothesized that the technique would work at 4 light years (and greater depending on knowledge on cloud abundance and a coronagraph design) meaning it could be tested on the recently discovered planet in the habitable zone of Proxima Centauri.  

The method was then tested empirically \citep{Doughty2016DetectingEarth} using the Galileo space probe data and first principles, in a similar methodology as~\citet{Sagan1993ASpacecraft}.  \citet{Sagan1993ASpacecraft} detected multiple stages of life on Earth, but they did not have a technique to distinguish between single and (non-technological) multicellular life on Earth.  \citet{Doughty2016DetectingEarth} used the Galileo space probe data but because the Galileo dataset had only a small change ($<2^\circ$) in phase angle (sun-satellite position), the observed anisotropy signal was small, and they could not detect multicellular life on Earth.  In contrast, in this paper, we propose to use to the POLDER satellite (Polarization and Directionality of Earth's reflectance) data to test this question.  This dataset gives global reflectance, directionality (BRDF), and polarization measurements at 20km resolution and phase angles of $>60^\circ$ \citep{Bicheron2000BidirectionalSpace}.  Therefore, we can create a view of Earth at different phase angles and determine empirically if, even scaling to a single pixel, we could distinguish between single and multicellular life on Earth.  

However, could the BRDF technique distinguish between abundant vertical structures like moon craters and abundant vegetation on an exoplanet?  Most such craters would in theory be eroded on a lifeless planet with weather and climate.  However, we test the BRDF of craters on Earth to understand how they cast shadows at different sun angles.    We took advantage of moon-like craters near our university that were created by the USGS in 1967 to help Apollo astronauts train by simulating different-sized lunar impact craters.  A total of 497 craters were made within two sites comprising 2,000 square feet.  We fly a UAV above a cratered landscape at different sun angles meant to replicate the moon landing site. 

\begin{figure*}[htp]
\centering
    \includegraphics[width=0.95\textwidth]{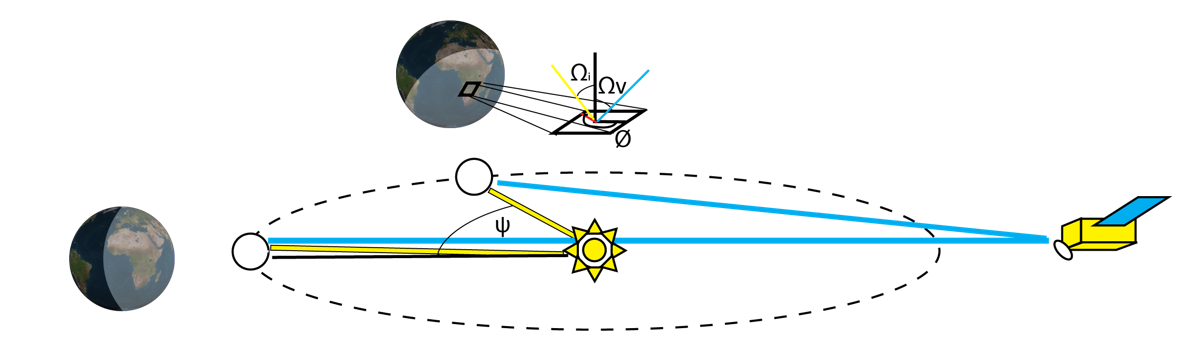}  
    \caption{Our conceptual design of a distant observer monitoring Earth and the change in backscattering as it revolves around the sun.   $\Theta$ is the azimuth angle, $\Omega$ i is the solar zenith angle, $\Omega$ v is the view angle, and $\Psi$ is the phase angle.}
   \label{fig:fig1}
\end{figure*}

We can also use detection of the red edge as corroborating evidence for the existence of vegetation. Our goal is to compare the reflectance properties at the red edge of plants with the BRDF or geometric optics, for example, the shape and arrangement of objects within a pixel that transmit or block light \citep{Torrance1967TheorySurfaces}, using Earth as an Exoplanet at various scales (Figure 1).  We propose to test this at the following scales: at the replica moon landing crater field, at the Amazon basin and the Sahara Desert, on all of Earth’s cloud free continental terrestrial surface and for the Earth as a whole.  We will then compare the phase function of the Earth as a single pixel to phase functions of other planets in the solar system.  We will compare Earth empirically (with POLDER data) and for Earth modelled with and without vegetation with a BRDF model \citep{Maignan2004BidirectionalSpot,Bacour2005VariabilityPOLDER} (Figure~\ref{fig:fig1}).

\section{Methods}
\label{sec:examples}

\subsection{Site information }
\begin{figure}[htbp]
\centering
\fbox{\includegraphics[width=.8\linewidth]{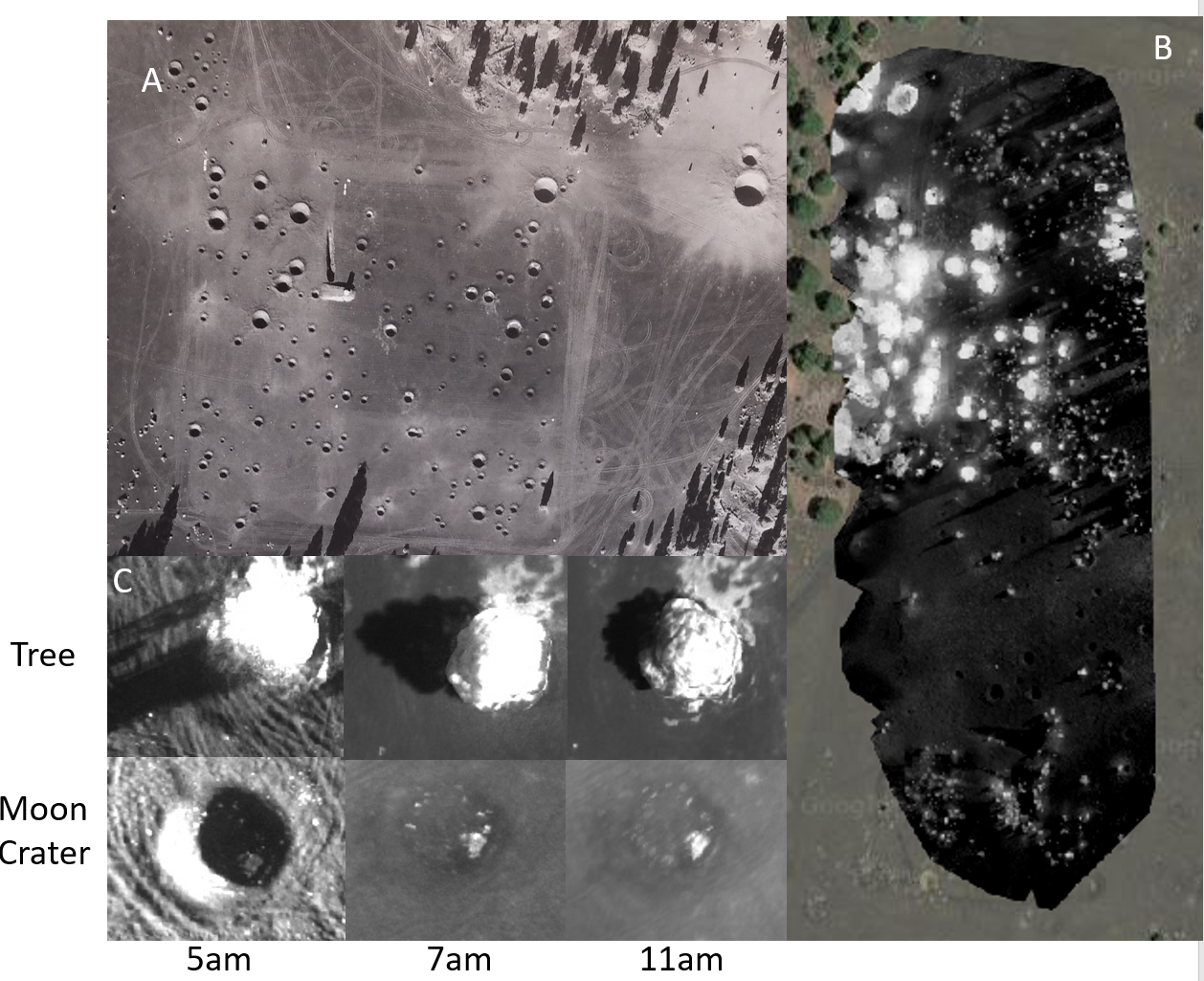}}
\caption{(A) The Apollo Astronaut training ground as originally photographed in 1967 (from USGS archives). (B) An example of UAV flyover measuring NDVI at 5am in 2018 with a current google earth image as a background image. (C) Closeups of two example regions of interest (tree and crater) at three different times of the day in the NIR (790 nm) band.  Note, the crater shadows visible at 5 am but not at later times while tree shadows are visible at all three times.}
\label{fig:fig2}
\end{figure}

To test NDVI and BRDF as biosignatures, we took advantage of an ``extraterrestrial landscape'' near our university that we call the replica moon landing crater site (35.30594920 lon, -111.50617530 lat).   Moon-like craters were created by the USGS in 1967 by digging holes and filling them with various amounts of explosives, which were detonated to simulate different-sized lunar impact craters.  The human-made craters range in size from 1.5--12 meters in diameter.  This area was chosen for the craters because of the basaltic cinders from an eruption of the Sunset Crater Volcano 950 years ago. After the explosions, the excavated lighter clay material spread out from the blast craters and across the fields, like ejecta from actual meteorite impacts. A total of 497 craters were made within two sites comprising 2,000 square feet (Figure~\ref{fig:fig2}).  

\subsection{UAV data acquisition }
We flew the Parrot Bluegrass (Parrot) UAV with 4 wavelengths (green 550 nm (40nm bandwidth (bw)), red – 660 nm(40nm bw), red edge 735nm (10 nm bw), and NIR 790nm (40 nm bw)) above the replica moon landing crater site described above.  We flew at various times to get different sun instrument angles (5:30, 7:30, 9:00 and 11:30 am) comparing three landscape types (bare ground, craters, and ponderosa pine trees).  The Parrot takes $\sim$200 photos at a height of 50m in each of the wavebands which are combined to form a map of $\sim$300 m\textsuperscript{2} (88 by 338m or 6ha) with a resolution of 4.7cm/pixel.  We use the program Pix4DCapture to plan the flight paths and Pix4Dmapper to orthomosaic the raw images into reflectance values (WGS 84 coordinate system).  This program created geotiffs for each band which we uploaded into the Google Earth Engine.  We used matlab (Mathworks) to further analyze this data.  

\subsection{Empirical Earth at different phase angles with POLDER data}
POLDER (Polarization and directionality of Earth's reflectance) gives global reflectance, directionality (BRDF), and polarization measurements~\citep{Bacour2005VariabilityPOLDER,Bicheron2000BidirectionalSpace}.  The ground size or resolution of a POLDER-measured pixel is 6x7 km$^2$ at nadir.   12 directional radiance measurements at each spectral band are taken for each point on Earth.  We downloaded data that capture the period from October 30, 1996 to February 28, 1997.  During that period, we chose 21 days interspersed within this broader period and aggregated data from those days (Specifically – Oct 30,31, Nov 1--6 and Dec 30--31 1996 and Jan 8, 9, 10, 11, 12, 14, 16, 17, 22, 23, 27 1997).  We also collected solar zenith angle (which is relative to the local zenith and may vary between 0$^\circ$ (sun at zenith) and approximately 80$^\circ$) and view zenith angle, (which is relative to the local zenith and may vary between 0$^\circ$ (POLDER at zenith) and approximately 75$^\circ$) (see Figure~\ref{fig:fig1} for an example of the geometries).  For each day, we subtracted the view zenith angle from the solar zenith angle (but we did not control for azimuth angle) to estimate phase angle for the wavelengths 565 nm (20 nm bandwidth) and 763 nm (10 nm bandwidth).  POLDER also has bands 670 nm and 865 nm, which are closer to traditional NDVI bands \citep{Masek2006} and have been used previously to characterize vegetation cover and BRDF responses \citep{Bacour2005VariabilityPOLDER}. However, these bands are also not ideal as they use polarized filters which are unlikely to be on future space telescopes.  Therefore, we use bands 670 nm and 865 nm in Figs S1-2 and table S1, but use 565 nm and 763 nm in the rest of the manuscript.  These two wavelengths were then used to create NDVI (Normalized difference vegetation index) according to the following equation:

$$NDVI = (763nm-565nm)/(763nm+565nm)$$
 
 We then created separate data maps for $<1^\circ$ phase angle ranges, then 1--3$^\circ$, then 3--6$^\circ$, 6--20$^\circ$, and 20--30$^\circ$.  We aggregated all available data for these five different phase angles and created cloud free land images of the Amazon basin, the Sahara Desert region, and all regions combined together.  We averaged these maps as a single pixel at the different phase angles to replicate what Earth might look like to a distant observer as it circles the sun at different phase angles. 

 \subsection{Modelled Earth at different phase angles with a BRDF model}
We used simulations of Earth with and without vegetation from Doughty et al 2010 at different phase angles.  In that paper, they used a semi-empirical BRDF model \citep{Bacour2005VariabilityPOLDER,Bicheron2000BidirectionalSpace}.  It combines a geometric kernel (F1), which models a flat Lambertian surface covered with randomly distributed spheroids with the same optical properties as soil \citep{Lucht2000AnModels}, with a volumetric kernel (F2), which models a theoretical turbid vegetation canopy with high leaf density \citep{Maignan2004BidirectionalSpot}.  They simulated global cloud cover with CAM 3.0; http://www.ccsm.ucar.edu/models/atm-cam) \citep{Collins2006TheCAM3}, and combined simulated cloud height (low, medium, and high) and total percent cover with albedo values for low, medium, and high clouds (strato-cumulous, alto-stratus, and cirrus) at several planetary phase angles \citep{Tinetti2006DetectabilityModel}.

\subsection{Other planets }
To compare the how Earth would look circling the sun at a distance to other planetary bodies, we digitized data from Sudarsky et al.~\citep{Sudarsky2005PhasePlanets} where they aggregated data for optical phase functions for Mars, Venus, the moon, and Uranus along with a Lambert model where radiation is scattered isotropically off a surface regardless of its angle of incidence \citep{Sudarsky2005PhasePlanets}. A classical phase function normalizes planetary albedo to 1 at a phase angle of 0$^\circ$. Data for Mars is originally from~\citet{Thorpe1977Viking1976}, for the Moon from~\citet{Lane1973MonochromaticDisk}, for Uranus from~\citet{Pollack1986EstimatesNeptune} and~\citet{Sudarsky2005PhasePlanets} does not state where the Venus data is originally from.  

We normalized all the datasets (Earth-POLDER, Earth no vegetation, Earth with vegetation, Mars, Venus, Uranus, and the moon) so that the albedo at phase angle of 0$^\circ$ was one.  We then subtracted these from a Lambert curve to highlight the impact of directional scattering from each of these bodies.

\section{Results}

The Apollo astronaut training ground offers a unique opportunity to compare NDVI and BRDF in an ``extraterrestrial landscape'' with trees.  In 1967, a flyover of the area early in the morning shows large shadows for both the craters and the local ponderosa pine trees (Figure~\ref{fig:fig2}(a)).  It is therefore conceivable that craters could replicate the shadows and BRDF is not a good multicellular life biosignature.  However, our UAV demonstrates why at later times of the day (at lower phase angles) the story changes.  Figure~\ref{fig:fig2}(b) shows our UAV NDVI image for the region at 5am.  The trees clearly have a higher NDVI and the craters still have shadows.  However, Figure~\ref{fig:fig2}(c) shows strong shadows with the craters at 5:30am but not at 9am and 11am.  In contrast, the trees show clear shadows at all times even towards noon (at lower phase angles).  This effect will change slightly with latitude \citep{Doughty2010DetectingPlanets}.   

\begin{figure}[htbp]
\centering
\fbox{\includegraphics[width=.8\linewidth]{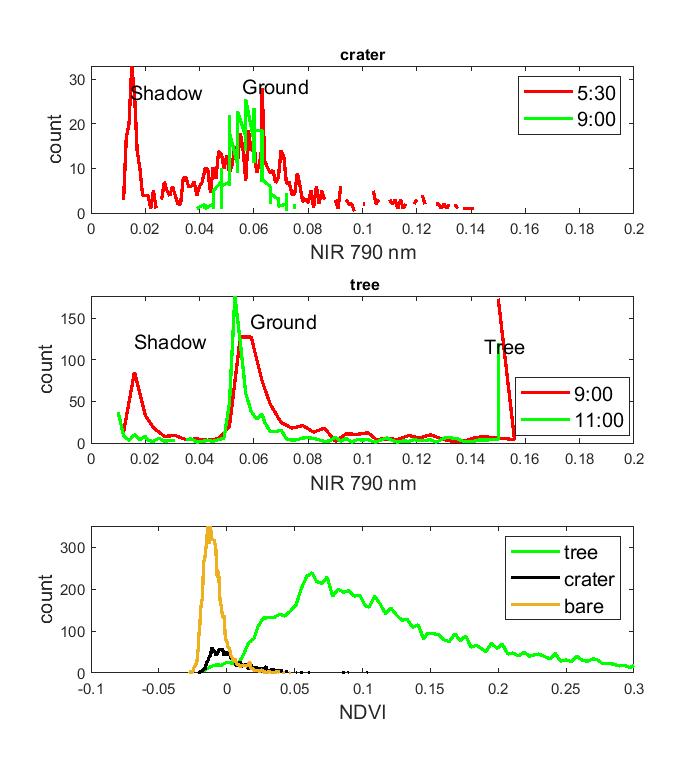}}
\caption{Histograms of NIR reflectance (790 nm) for (top) craters and (middle) trees at different times of the day (5am and 9am for craters, 9am and 11am for trees).  For clarity, we aggregate all tree reflectance pixels greater than 0.15 to 0.15.  (bottom) NDVI for trees (green), craters (black), and bare ground (blue) at 11am.}
\label{fig:fig3}
\end{figure}

We can quantify these qualitative observations with our UAV collected reflectance data.  Figure~\ref{fig:fig3} shows the reflectance histograms for trees and craters in the NIR (790 nm) at different times of the day.  Because the UAV flew overhead, the daytimes correspond with high (5am), medium (9am) and low phase angles (11am).  In Figure~\ref{fig:fig3}(a), at 5:30 am the histogram of the crater shows a strong shadow peak at $\sim$0.01 reflectance and another reflectance peak at $\sim$0.05 reflectance.  However, by 9am the shadow peak disappears and there is only the ground reflectance peak at $\sim$0.05 reflectance.   In Figure~\ref{fig:fig3}(b), at 9am there are reflectance peaks for shadows at $\sim$0.01 reflectance, at the ground at $\sim$0.05 reflectance, and for the tree canopy which was scattered but for clarity we reduced to 0.15 reflectance.  At 11am, there are similar peaks, but with a small number of shadow pixels at 0.01 reflectance as expected.   The difference between the peak brightness at 0$^\circ$ phase angle and reduced brightness at higher phase angles is our hypothesized ``multicellular life biosignature''.

NDVI showed different reflectance peaks for trees than for bare ground and craters.  The ``tree'' NDVI signal included shadows and bare ground which reduced the overall NDVI signal.  However, even with the mixed signal, NDVI also showed a clear signal that could distinguish between the three areas with NDVI’s median histogram of 0.06 for the trees and $\sim$0 for both the crater and bare ground (Figure~\ref{fig:fig3}(c)).    Was the NDVI or BRDF signal greater?  For example, a typical region of interest with 50 percent tree cover, 50 percent ground at 9am might have 25 percent of the ground covered in shadow.  At 9am, our scene might have an NIR reflectance of 0.09 (0.15*0.5 (tree)+0.01*0.25 (shadow)+0.05*0.25 (ground)) while at noon, as the shadows are masked, it would change to 0.10 (0.15*0.5 (tree)+0.05*0.50 (ground)).  This is a relatively small change of 0.01.  We have shown that moon craters would not show this change and the 0.01 signal is the ``multicellular life biosignature''.  However, the NDVI signal of $\sim$0.06 is clearly larger.

Next, we scaled up to the regional and global scale with POLDER data.  We first created cloud free terrestrial maps of Earth at 5 different phase angles.  We found that the $<1^\circ$ phase angle contained many regional blank areas, especially tropical regions with great cloud cover, and we did not include it in our final analysis.  We discuss this more in the discussion section.  Therefore, we focused on the phase angle ranges of 1--3$^\circ$, 3--6$^\circ$, 6--20$^\circ$, and 20--30$^\circ$.  Averaging over 21 days gave sufficient cloud free images to create maps for most of the planet.  There were still gaps in our coverage, both at high latitudes, where POLDER did not cover, and in parts of the tropics where clouds were very abundant.

\begin{table}[htbp]
\centering
\caption{\bf Absolute change of reflectance (between 1--3$^\circ$ phase angle and 20--30$^\circ$ phase angle) for band 763 nm, NDVI and the percent change for band 763 nm for the Amazon, Sahara, all land and the world.}
\begin{tabular}{ccccc}
\hline
 & Amazon & Sahara & All land & world\\
\hline
$763 nm$ & $0.016$ & $0.007$ & $0.015$ & $0.012$\\
$NDVI$ & $0.055$ & $0.009$ & $0.043$ & $0.033$\\
$per763 nm$ & $8.5$ & $3.8$ & $10.9$ & $8.2$\\
\hline
\end{tabular}
  \label{tab:shapefunctions}
\end{table}

These cloud free images allowed us to compare two multicellular life endmembers – the Amazon basin, with abundant tree cover, and the Sahara Desert, with very few trees.  In Figure~\ref{fig:fig7}(a), we show the average reflectance for these two regions at both 565 and 763nm at several different phase angles.  The changes were smaller than we had hypothesized with our BRDF model possibly because we missed the large change between 0--1$^\circ$ phase angle.  At 763nm between phase angle 1--3$^\circ$ and 20--30$^\circ$ there was a difference of 0.016 reflectance units or $\sim$9 percent for the Amazon versus 0.007 reflectance units or $\sim$4 percent for the Sahara (Table 1).  There were only minor changes for the Sahara or for the Amazon at 565nm. We show results using polarized bands 670 and 865nm (Figs S1-2 and Table S1) and show an overall larger NDVI signal, but similar changes in reflectance at different phase angles. The supplemental data demonstrate that our results are robust for all POLDER wavelengths tested.  The slight improvement at bands 670 and 865 nm is most likely due to less atmospheric interference (the O2-A band interferes at 763 nm and aerosols interfere at 565 nm) and not the polarized filter. Bands near 670 and 865 nm would therefore be our choice for future space missions.  

\begin{figure}[htbp]
\centering
\fbox{\includegraphics[width=.8\linewidth]{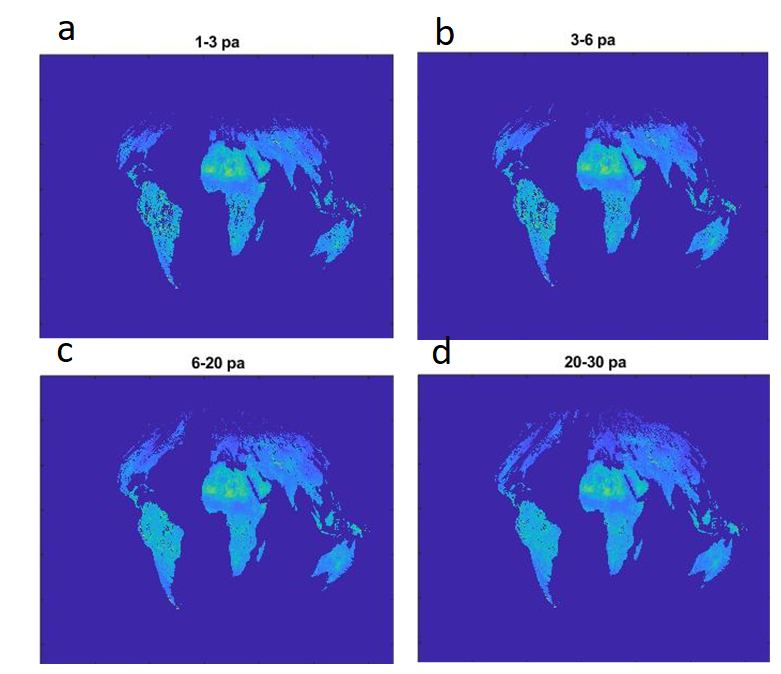}}
\caption{Cloud free terrestrial reflectance at 763nm from POLDER at the phase angle (pa) ranges (a) 1--3$^\circ$, (b) 3--6$^\circ$, (c) 6--20$^\circ$, and (d) 20--30$^\circ$ aggregated and averaged from the 21-day period described in the methods.}
\label{fig:fig4}
\end{figure}

\begin{figure}[htbp]
\centering
\fbox{\includegraphics[width=.8\linewidth]{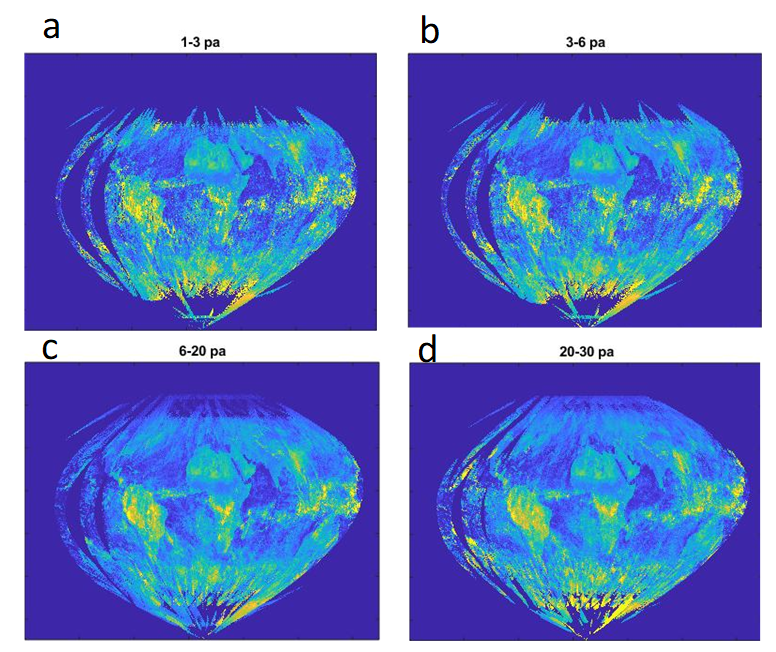}}
\caption{All pixels (including ocean and clouds) reflectance at 763nm from POLDER at the phase angles (a) 1--3$^\circ$, (b) 3--6$^\circ$, (c) 6--20$^\circ$, and (d) 20--30$^\circ$ aggregated and averaged from the 21-day period described in the methods.}
\label{fig:fig5}
\end{figure}

\begin{figure}[htbp]
\centering
\fbox{\includegraphics[width=.8\linewidth]{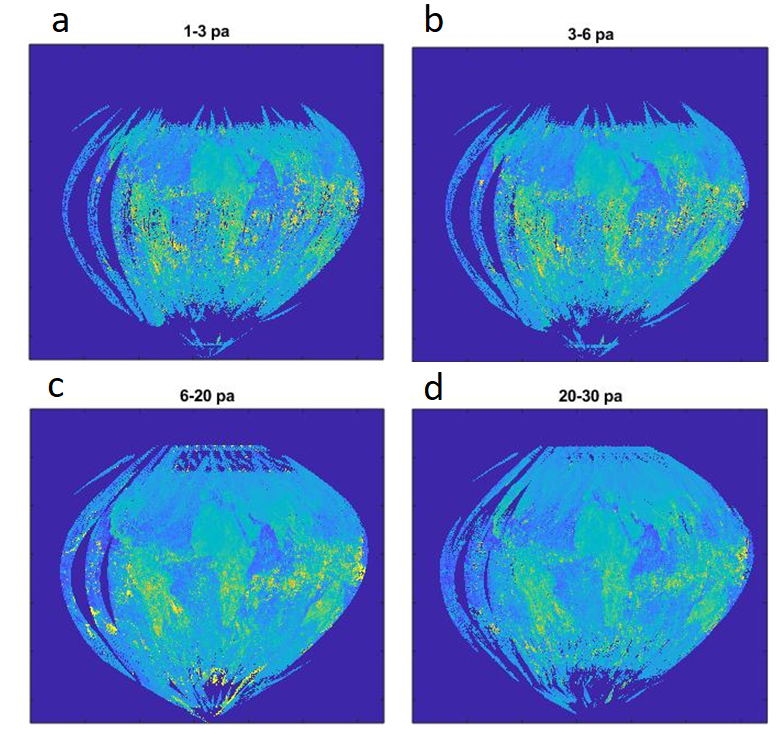}}
\caption{All (including ocean and clouds) NDVI pixels from POLDER at the phase angles (a) 1--3$^\circ$, (b) 3--6$^\circ$, (c) 6--20$^\circ$, and (d) 20--30$^\circ$ aggregated and averaged from the 21 day period described in the methods.}
\label{fig:fig6}
\end{figure}

\begin{figure}[htbp]
\centering
\fbox{\includegraphics[width=.8\linewidth]{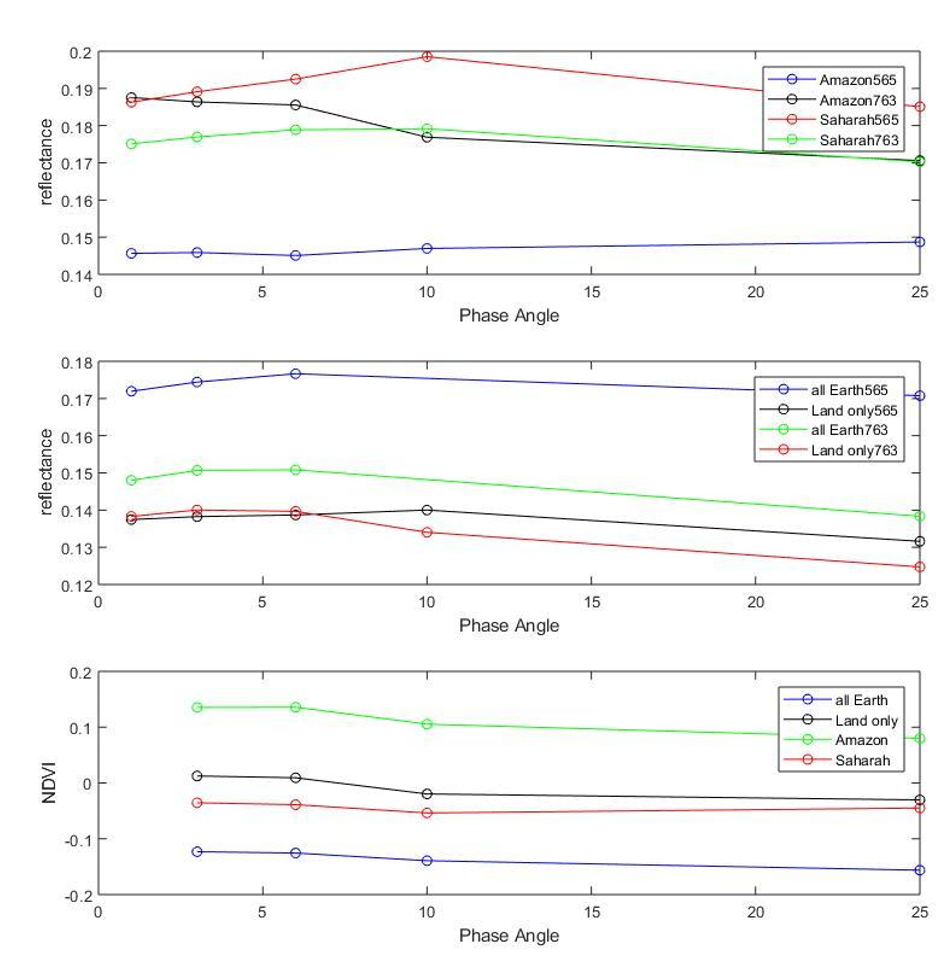}}
\caption{(top) Averaged reflectance at different phase angles at different wavelengths (565 nm and 763 nm) for the Amazon region and the Sahara region.  (middle) Averaged reflectance at different phase angles for all Earth and all terrestrial land at different wavelengths (565 nm and 763 nm).  (bottom) Averaged NDVI at different phase angles for a cloud covered Earth (red), all terrestrial land (black), the Amazon region (green), and the Sahara region (blue). }
\label{fig:fig7}
\end{figure}

We next created a global view of Earth (including land, clouds and oceans) at the different phase angles (Figure~\ref{fig:fig5}) and a NDVI of the entire Earth at different phase angles (Figure~\ref{fig:fig6}).  In Figure~\ref{fig:fig7}(b)~and~(c), we average Figure~\ref{fig:fig4},~\ref{fig:fig5},~and~\ref{fig:fig6} as a single pixel at the different phase angles.  As a single pixel, at 565nm, there are only minor reflectance changes between phase angle 1--3$^\circ$ and 20--30$^\circ$.  However, at 763 nm, the land only had reflectance changes $\sim$0.015 or $\sim$12 percent and the whole world had a slightly smaller change of 0.011 or $\sim$8 percent (Table 1).  We also compared averaged NDVI for the Amazon, the Sahara, all land and the averaged planet to combine information on the red edge with BRDF.  As expected, the Amazon had the highest NDVI followed by all land, the Sahara and the whole world.  The decrease in NDVI across phase angles was similar (0.06) for the Amazon, the land (0.04) and the world (0.03) but stayed flat for the Sahara (0.01) (Table 1).   

Finally, we combined information from POLDER for Earth and compared this to measured estimates for other planetary bodies such as Mars, Venus, the Moon, and Uranus.  We also added estimates of a Lambert body (a body with perfect isotropic reflectance) and modelled Earth with and without vegetation \citep{Doughty2010DetectingPlanets}.   All planetary bodies have very different albedos, but for comparison purposes, we standardized the average albedo to 1 at a phase angle of 0.  We initially hypothesized that Earth would have a phase function between Mars and Venus (with both POLDER and the vegetation model in agreement).  In other words, Earth might be a partially cloudy planet with some directional reflectance.  However, our modeled estimates of Earth, with and without vegetation showed similar directional reflectance to Mars but our empirical results using POLDER data showed Earth was more similar to Venus (Figure~\ref{fig:fig8}).

\section{Discussion}

\begin{figure}[htbp]
\centering
\fbox{\includegraphics[width=.8\linewidth]{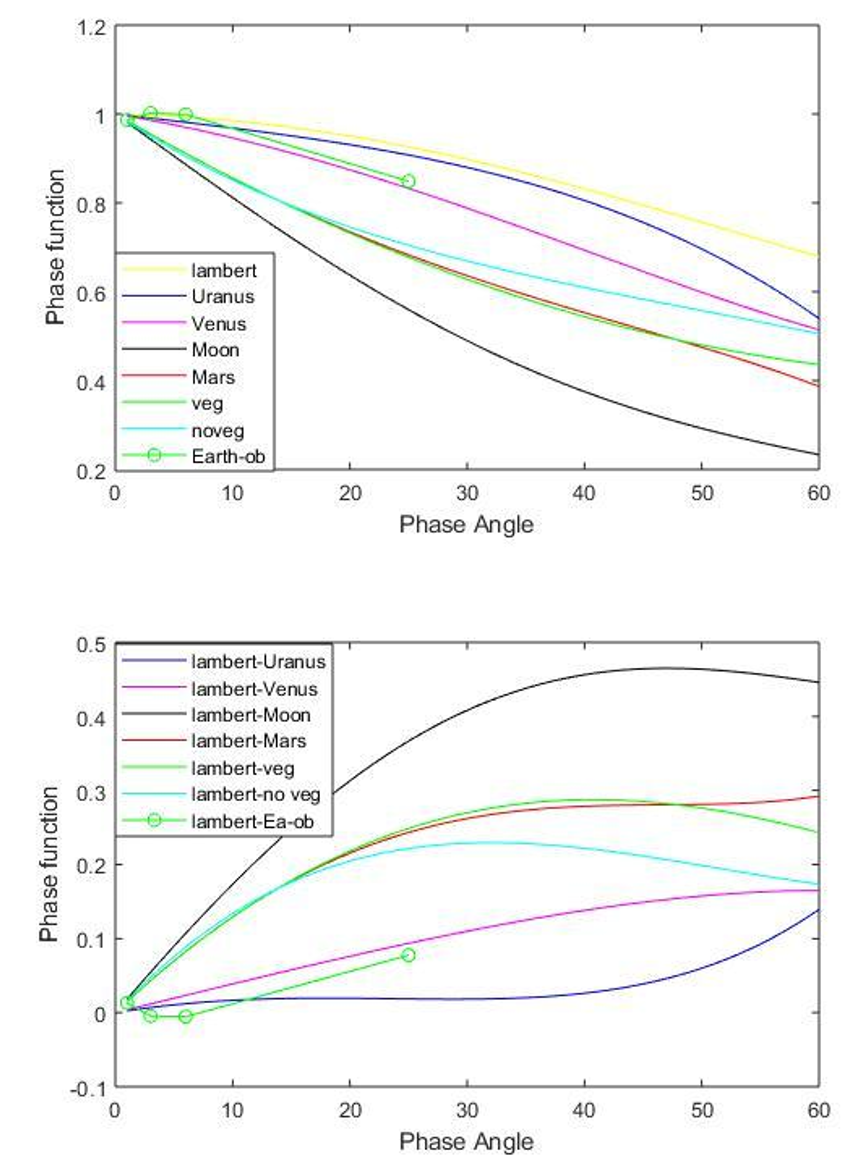}}
\caption{(Top) The phase function for several solar system objects (from~\citet{Sudarsky2005PhasePlanets}), Earth with and without vegetation structure (from~\citet{Doughty2010DetectingPlanets}) and empirically calculated for a cloud covered Earth with POLDER data from this paper.  The phase function normalizes for albedo by forcing albedo to one at a phase angle of 0$^\circ$.  We also show a lambert model from~\citet{Sudarsky2005PhasePlanets} which assumes an object that scatters light perfectly isotopically.  In the bottom figure, we show the same data but subtract the Lambert curve to more clearly show backscattering differences. }
\label{fig:fig8}
\end{figure}

Why there was a large divergence between our modelled results of Earth at different phase angles and our empirical ones?  To review, modelled Earth’s reflectance at different phase angles is similar to Mars while empirical POLDER data of Earth’s reflectance at different phase angles are similar to Venus (Figure~\ref{fig:fig8}).  We hypothesize that both the model and empirical data have issues that make them not align.  For instance, our model uses the best vegetation BRDF model, but it did not have a good BRDF model for other components of the Earth, such as oceans, clouds and atmosphere.  Therefore, it likely missed key components of atmospheric scattering and cloud directional reflectance.  In contrast, we hypothesize that there were also issues with the empirical data because by excluding our phase angle data of $<1^\circ$ degree in our empirical analysis, we missed the largest change in BRDF.  Our BRDF model suggests the largest change in reflectance from vegetation will be between phase angles of 0-1$^\circ$ and 1--3$^\circ$.  Therefore, by missing this peak, and showing little change $<10^\circ$, our phase curve is more like an isotropic body like Venus.  

Mars and the moon both have greater backscattering than Earth.  For solid bodies with thin atmospheres like Mars, previous work has shown that backscattering can be significant \citep{Thorpe1977Viking1976}.  This is because Mars (currently) has no liquid water to erode and smooth its rough edges.  Our phase curve (Figure~\ref{fig:fig8}), shows that the moon has even stronger backscattering than Mars, which is initially surprising \citep{Lane1973MonochromaticDisk}.  However, this is due to a phenomenon called coherent backscatter which occurs on very dry soils where particles have a diameter that is similar to the wavelength of the photon used to view them \citep{Hapke1993TheBackscatter}.  A planet with climate like Earth does not exhibit coherent backscatter, even in dry areas, such as deserts, because the particle sizes are too big (generally between 0.05 to 2mm) at 800 nm or less \citep{Tarbuck2008EarthGeology}.   Therefore, Earth shows less backscattering than Mars or the moon because of the presence of abundant isotropic clouds.  The presence of craters on the moon and Mars also affects backscattering.  At low phase angles the BRDF of craters is substantially different than that of trees (Figures~\ref{fig:fig2}~and~\ref{fig:fig3}).  Earth has few craters due to abundant erosion caused by climate.  It is interesting to note the large amount of erosion of the craters at the replica moon landing site that has already occurred due to weather and climate in the 50 years since the craters were first formed.  

In contrast, Venus and Uranus have scattering more similar to Lambert scattering where radiation is scattered isotropically off a surface. Lambert scattering is a good approximation for objects such as Uranus \citep{Pollack1986EstimatesNeptune}, and to a lesser extent Venus \citep{Sudarsky2005PhasePlanets}.  Surprisingly, our empirically derived phase function for Earth was less steep than either Venus or Uranus (Figure~\ref{fig:fig8}).  This is surprising because Earth has many strong backscattering surfaces like trees.  We hypothesize that this is due to excluding our phase angle data of $<1^\circ$ in our empirical analysis.

To improve our future empirical analysis, we need to better capture low phase angles.  With the POLDER data, averaging for phase angles of 1 degree or less was inherently more patchy because it was averaging over a smaller dataset.  Key regions, like Amazonia were missing because of high cloud cover.  In fact, the cloudier terrestrial areas, and the regions less represented at $<1^\circ$ phase angle, were those most likely to have abundant tree cover (like Amazonia).  For this reason, we were not confident including our maps of $<1^\circ$ phase angle.  POLDER was only available for a few months during 1996-1997 and it is currently the only satellite of its kind to capture the Earth at all phase angles.  Capturing planets at low phase angles will also be a problem with any viewing of an exoplanet because it could be washed out by the light of its star, even with the most advanced coronagraph design \citep{Guyon2006TheoreticalCoronagraphs}.   However, in theory, we could observe the planet during continuous rotation cycles which could increase the amount of data available to analyze the exoplanet for vegetation structure.

To improve our modelling analysis, we need to better model the BRDF of non-vegetated surfaces.  We used a state of the art BRDF model for vegetation \citep{Bicheron2000BidirectionalSpace,Bacour2005VariabilityPOLDER}, but only averaged BRDF values for clouds, atmosphere and oceans.   With this improved model, how do we envision using the model in the future to distinguish a planet with multicellular life versus just single cellular life?  We could create a model of an exoplanet based on the exoplanet’s size, density, cloud cover, distance to star, and the star’s irradiance.  For instance, let us imagine we had the proper technology and coronagraph to observe the 1.3 Earth mass planet only $\sim$4 light years from Earth within the habitable zone of the red dwarf Proxima Centauri \citep{Anglada-Escude2016ACentauri}.  We would then create three versions of the model, first a relatively smooth, eroded, planetary surface, one covered with single cellular slime exhibiting NDVI and one with 3D vegetation structure.  We would look for evidence of which model better fit observations of the exoplanet over years.  False positives caused by instrument error or intermittent events such as volcanic activity or changing cloud cover could be determined by observing the planet during continuous rotation cycles. Multicellular life would continuously demonstrate the BRDF signal, while other causes would demonstrate it only intermittently.  In practice, this will be difficult with the next generation potential space telescopes for directly imaging exoplanets such as HabEx and LUVOIR.  These are predicted to have 10--20 signal to noise ratio (SNR) for exoplanet spectroscopy but if an exciting target were to be discovered, more telescope time could increase this to $\sim$20--100 SNRs.  

In our work, we assumed that the observer is in an approximate plane with the planet’s orbit and its star.  However, Proxima Centauri b is now not assumed to transit its star \citep{Jenkins2019},  and therefore observing Proxima Centauri b at small phase angles (~< 3 degrees) is likely impossible with any future technology.    However, the assumption of the observer being in an approximate plane with the planet’s orbit will be valid for Earth-sized planets detected by the transit method (such as by the Transiting Exoplanet Survey Satellite (TESS) or ground-based surveys).  The BRDF technique would require additional geometric calculations for planets not meeting this assumption.  

It will be important to understand potential false positives if we are to have confidence in this approach in the future.  For instance, Livengood et al.2011) found that the NDVI of the moon’s disk-average is greater than the Earth’s disk-average \citep{Livengood2011}. Since the moon obviously has no plants, this exemplifies the need to carefully think through all potential ways the BRDF signal could be created without vegetation.  For instance, we could mistake stromatolites, which are some of the earliest evidence for singlecellular life on Earth \citep{Walter1980}, for trees due to geometric similarities.  However, generally, microbes do not tend to display a strong red edge, so one possibility is a red edge filter.  Stromatolites also tend to be in shallow water, a rare environment for trees.  Since water has a vastly different reflectance spectrum than dry ground, this could be a second filter.  Therefore, a stromatolite covered planet could replicate some of the geometry of trees (although the geometry itself is also much different), would have a much lower average albedo in NIR, and would likely not have a strong red edge. 

In the more distant future many such issues may be resolved with the advent of new technologies and spatially resolved imaging of Earth-size exoplanets may be possible. Conceptual designs include the Exo-Earth mapper \citep{Kouveliotou2013} and the Solar Gravity Lens Project \citep{Turyshev2018}.  While these concepts are quite ambitious, they are far more plausible than interstellar travel and would provide an opportunity to search for the geometric signatures of multicellularity as outlined in this manuscript.

\section{Conclusions and future directions}

Overall, in theory, BRDF could distinguish between multicellular and single cellular life on exoplanets, but we have recognized issues with both our models and our empirical observations that must be improved before this technique could be used with confidence.  The easiest short-term step is to improve the modelling by combining the various BRDF models.  Further empirical validation will be more challenging as POLDER is a unique satellite. Here we demonstrate that BRDF is challenging to detect and will be a smaller signal than NDVI, which has already proven to be challenging to detect with Earth as an exoplanet \citep{MontanesRodriguez2006VegetationPlanets}.  Should this line of research therefore be abandoned?  Theoretically, it could still work and since we are not aware of other techniques to distinguish an exoplanet with multicellular life, we believe further work should still continue.

\ack[Acknowledgement]{This project was funded by NASA’s Habitable World’s program with the project name: ``Testing methods to detect 3D vegetation structure on exoplanets'' (16-HW16-2-0025). }

\bibliography{astro}

% \begin{thebibliography}{}

% \bibitem[Adade \textit{et~al.}(2003)]{Chen2001}
% \textbf{Adade CM, de Castro SL and Soares MJ} (2007) Ultrastructural localization of
% \textit{Trypanosoma cruzi} lysosomes by aryl sulphatase
% cytochemistry. \textit{Micron} \textbf{38}, 252--256.

% \bibitem[Bayer-Santos \textit{et~al.}(2002)]{Chen1995}
% \textbf{Bayer-Santos E, Aguilar-Bonavides C, Rodrigues SP, Cordero
% EM, Marques AF, Varela-Ramirez A, Choi H, Yoshida N, da Silveira JF
% and Almeida IC} (2013) Proteomic analysis of \textit{Trypanosoma
% cruzi}
% secretome: characterization of two populations of extracellular
% vesicles and soluble pro-teins. \textit{Journal of Proteome Research}
% \textbf{12}, 883--897.
% \end{thebibliography}

%%%%%%%%%%%% Biography text %%%%%%%%%%%%%%%%%%%%%%%%%%

% \aubio{author_photo.eps}{\textbf{First A. Author} received a degree in physics from the University
% of A in 1998 and received his Ph.D. degree in communication engineering in
% 2002. He now holds a research chair at the B institute. His main research
% interests are design and optimization of high power microwave power
% amplifiers.}

% \aubio{author_photo.eps}{\textbf{Second B. Author} received her Diploma in 1997 from the University of
% C and received her Ph.D. at the D University in 2001. She became a full
% university professor in 2009. She is currently serving as chair of the E
% project and is active in the study of Si power electronics.}

\end{document}